\documentstyle[pre,aps,multicol,epsfig]{revtex}

\begin{document}
\title{Solitons in Triangular and Honeycomb Dynamical Lattices with the Cubic
Nonlinearity}
\author{P.G. Kevrekidis$^{1}$, B.A. Malomed$^2$ and Yu.B.
Gaididei$^3$}
\address{
$^1$ Department of Mathematics and Statistics, University of Massachusetts,
Amherst MA 01003-4515, USA \\
$^2$ Department of Inderdisciplinary Studies,
Faculty of Engineering, 
Tel Aviv University, Tel Aviv 69978, Israel \\
$^3$ N.N. Bogolyubov Institute for Theoretical
Physics, 03143 Kiev, Ukraine}
%\date{today}
\maketitle

\begin{abstract}
We study the existence and stability of localized states in the discrete
nonlinear Schr{\"o}dinger equation (DNLS) on two-dimensional non-square
lattices. The model includes both the nearest-neighbor and long-range
interactions. For the fundamental strongly localized soliton, the results
depend on the coordination number, i.e., on the particular type of the
lattice. The long-range interactions additionally destabilize the discrete
soliton, or make it more stable, if the sign of the interaction is,
respectively, the same as or opposite to the sign of the short-range
interaction. We also explore more complicated solutions, such as twisted
localized modes (TLM's) and solutions carrying multiple topological charge
(vortices) that are specific to the triangular and honeycomb lattices. In
the cases when such vortices are unstable, direct simulations demonstrate
that they turn into zero-vorticity fundamental solitons.
\end{abstract}

\date{today}

\section{Introduction}

In the past decade, energy self-localization in nonlinear dynamical
lattices, leading to the formation of soliton-like {\it intrinsic localized
modes} (ILMs), has become a topic of intense theoretical and experimental
research. Much of this work has already been summarized in several reviews 
\cite{BrKi1,Aub1,FlWi1,FlMk1,HeTsi1,KevRev}. It was proposed that this
mechanism would be relevant to a number of effects such as nonexponential
energy relaxation in solids \cite{EiSc1}, local denaturation of the DNA
double strand \cite{PeyBi1,DPB1,DPB2,DFrev1}, behavior of amorphous
materials \cite{KoAu1,KoAu2,KoAu3}, propagation of light beams in coupled
optical waveguides \cite{Jen,CJ,Ac1} or the
 self-trapping of vibrational energy in
proteins \cite{ChrSc1}, among others.
 ILMs also have potential significance in some
crystals, like acetanilide and related organics \cite{eilbeck,kalosakas}.
The theoretical efforts were complemented by a number of important
experimental works suggesting the presence and importance of the ILMs in
magnetic \cite{Sievers} and complex electronic materials \cite{Swan}, DNA
denaturation \cite{CaGia}, as well as in coupled optical waveguide arrays 
\cite{Eis1,Mor1} and Josephson ladders \cite{Trias,flach1}.

A ubiquitous model system for the study of ILMs is the discrete nonlinear
Schr{\"{o}}dinger (DNLS) equation (see e.g., the review \cite{KevRev} and
references therein). Within the framework of this model and, more generally,
for Klein-Gordon lattices, it has recently been recognized that physically
realistic setups require consideration of the ILM dynamics in higher spatial
dimensions \cite{KRB1,KMB,others,BUR,flach,FLA,AJS,KRB5,GCM1}. In the most
straightforward two-dimensional (2D) case, almost all of these studies, with
the exception of Refs. \cite{Eil1,Eil2} were performed for square lattices.
However, it was stressed in Ref. \cite{Eil1,Eil2} that non-square geometries
may be relevant to a variety of applications, ranging from the explanation
of dark lines in natural crystals of muscovite mica, to sputtering (ejection
of atoms from a crystal surface bombarded by high-energy particles), and,
potentially, even to high-temperature superconductivity in layered cuprates.
%%such as YBa$_{2}$Cu$_{3}$O$_{7}$, 
%%Y$_{2}$Ba$_{4}$Cu$_{6}$O$_{13}$ or La$_{2}$CuO$_{4}$. 
Besides that, it has been well recognized that triangular (TA)
and hexagonal (or {\it honeycomb}, HC) lattices are relevant substrate
structures in a number of chemical systems \cite{atencio} and, especially,
in photonic band-gap (PBG) crystals \cite{smir,lyngby}. Notice that, in the
context of the PBG crystals, the relevance of nonlinear effects has been
recently highlighted for a square diatomic lattice \cite{kiv}.

The above discussion suggests the relevance of a systematic study of ILMs in
the paradigm DNLS model for TA and HC lattices. The aim of the present work
is to address this issue (including the stability of the ILM solutions), for
the 2D lattices with both short-range and long-range interactions. In
section II we discuss the effects of the non-square lattice geometry on the
fundamental ILM state (the one centered on a lattice site), and then explore
effects of long-range interactions on this state. In section III, we expand
our considerations to other classes of solutions, which are either more
general ones, such as twisted modes, which are also known in square
lattices, or represent states that are specific to the TA and HC structures,
viz., discrete vortices. We identify stable fundamental vortices in the TA
and HC lattices  with vorticity (spin) $S=3$ and $S=5$, and with
the hexagonal and honeycomb shape, respectively. Additionally, a triangular
vortex is found in the HC lattice, but it is always unstable. 
%In comparison, it is relevant to mention that only the usual
%vortices with $S=1$ may be stable in square lattices \cite{DV}.

\section{Fundamental intrinsic localized modes}

\subsection{The model}

\begin{figure}[tbp]
\epsfxsize=10cm
\centerline{\epsffile{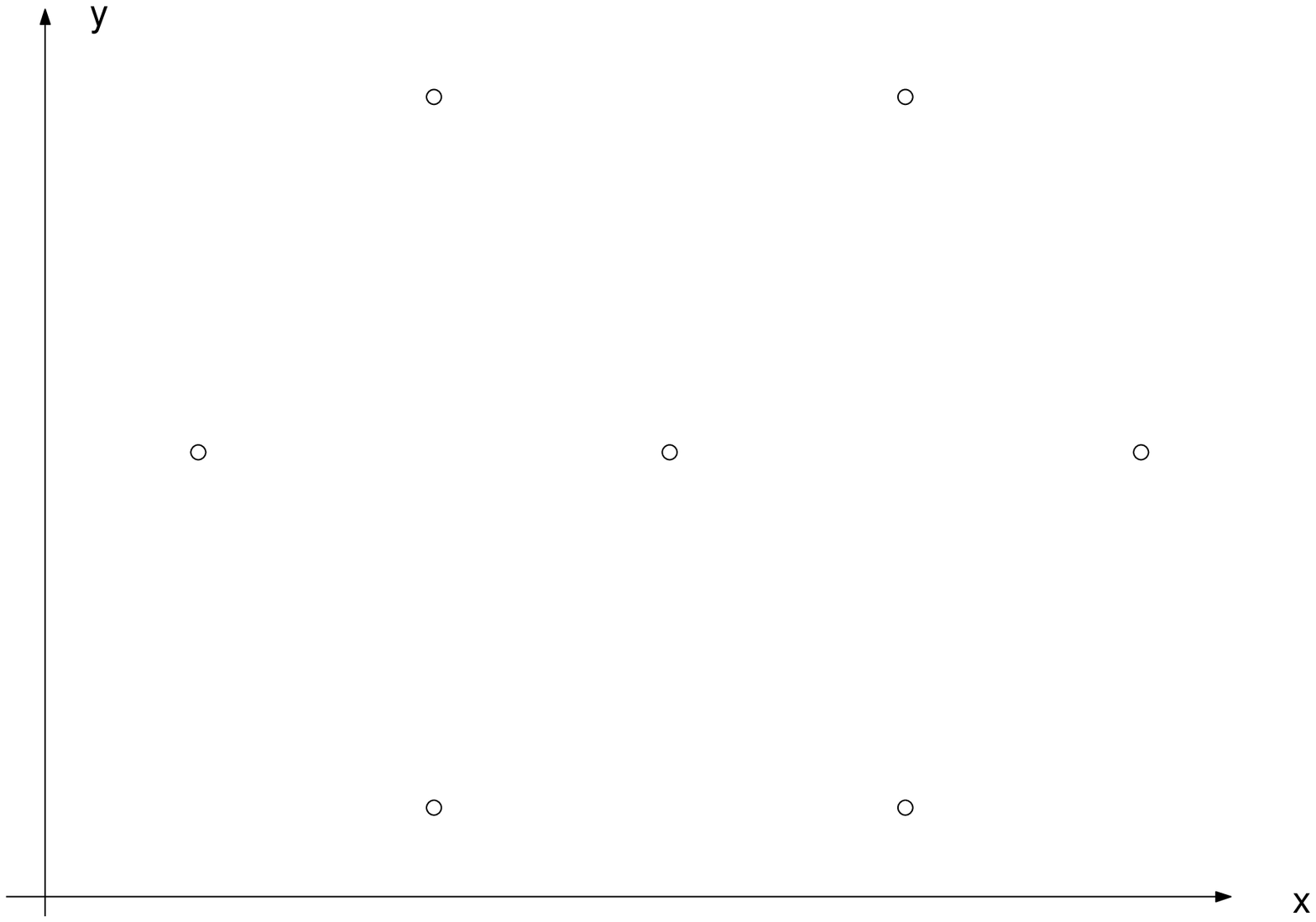}}
\epsfxsize=10cm
\centerline{\epsffile{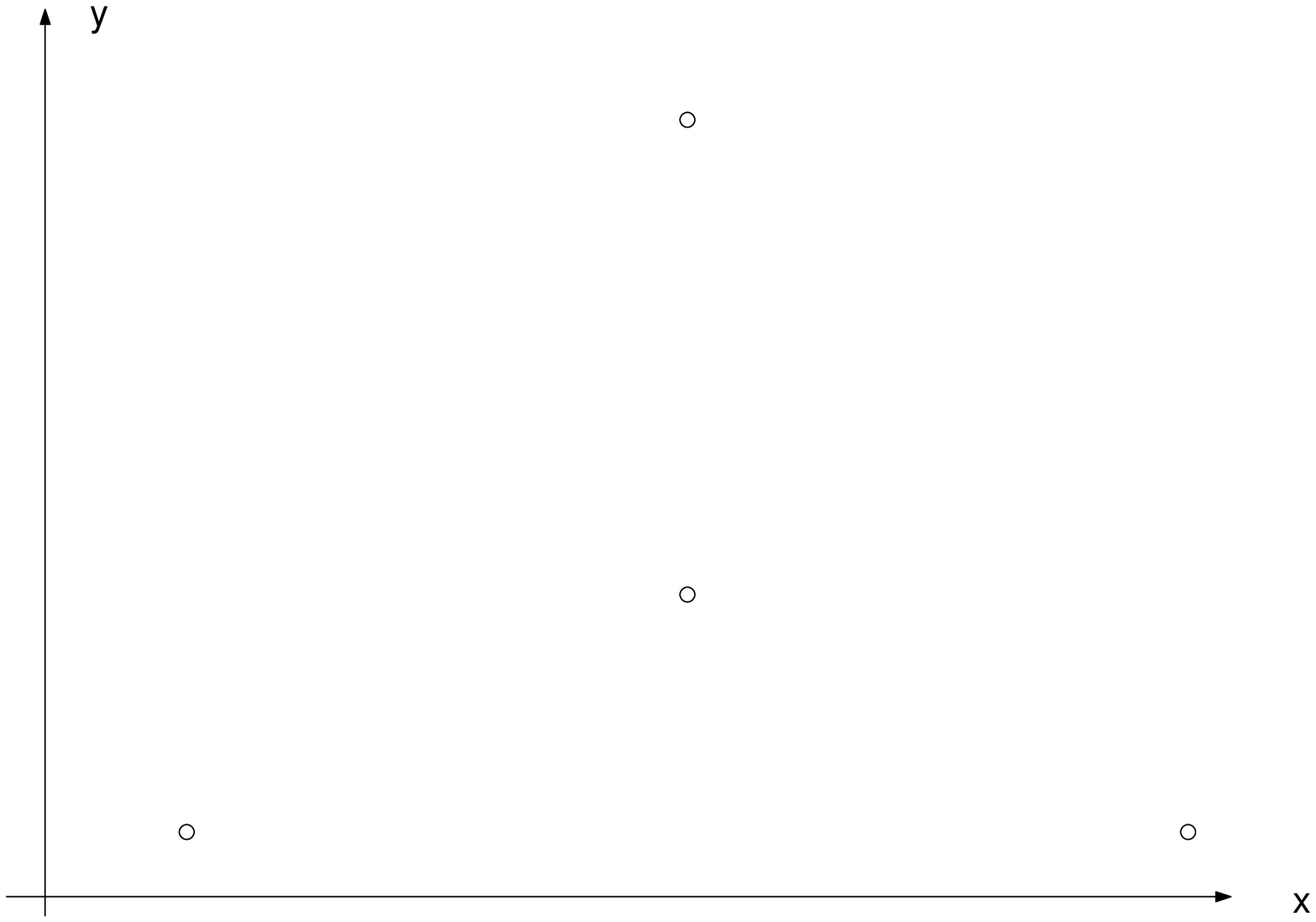}}
\caption{The top panel shows a cell of the triangular configuration
incorporating six nearest neighbors of a given site. Similarly, one of the
two possible configurations of neighbors in the honeycomb lattice is shown
in the bottom panel. An alternative possibility in the latter case involves
one neighbor along the negative y-axis and two neighbors along directions at 
$\pm \protect\pi/3$ angles with the positive y-axis (if we place the central
site at the origin of the coordinate system). }
\label{hfig1}
\end{figure}

In this work we consider the two-dimensional DNLS equation with the on-site
cubic nonlinearity, 
\begin{equation}
i\dot{\psi}_{nm}=-C\sum_{\left\langle n^{\prime },m^{\prime }\right\rangle
}\psi _{n^{\prime }m^{\prime }}+kC\psi _{nm}-|\psi _{nm}|^{2}\psi
_{nm}-\sum_{n^{\prime },m^{\prime }}{\bf K}\left( h\sqrt{(n-n^{\prime
})^{2}+(m-m^{\prime })^{2}}\right) \psi _{n^{\prime }m^{\prime }}\,,
\label{heq1}
\end{equation}
The subscripts ($n,m$) attached to the complex (envelope) field $\psi $ are
two discrete spatial coordinates, $C$ is the constant of the linear coupling
between nearest-neighbor sites, the summation over which is denoted by $
\left\langle ...\right\rangle $, and $k$ is the coordination number (i.e.,
the number of the nearest neighbors), which takes the values $k=6$ for the
TA lattice (see the left panel of Fig. \ref{hfig1}), $k=4$ for the square
lattice, and $k=3$ for the HC one (see the right panel of Fig. \ref{hfig1}).
The function ${\bf K}$ represents a kernel of the long-range linear
coupling, and $h\equiv 1/\sqrt{C}$ is the lattice spacing.

It is worth noting here that the TA network is a simple Bravais lattice with
the coordinates of the grid nodes $x_{nm}=h(n+m/2)$ and $y_{nm}=\sqrt{3}hm/2$
(see also Ref. \cite{smir}). The same is true for the most commonly used
square lattice, which has $x_{nm}=nh$, $y_{nm}=mh$, but {\em not} for the HC
structure, a simple representation of which (for $h=1$) is $x_{nm}=\sqrt{3}m$
, $y_{nm}=(1/4)\left[ 6n-5+(-1)^{n+m}\right] $. More information on the
latter structure (which also represents, for instance, the arrangement of
carbon atoms in a layer of graphite) and its symmetries can be found in Ref. 
\cite{indiana}.

First, we will look for ILM solutions of the nearest-neighbor version of the
model, setting ${\bf K}\equiv 0$. Stationary solutions with a frequency $
\Lambda $ are sought for in the ordinary form (see e.g., Ref. \cite{KevRev}), 
\begin{equation}
\psi _{nm}=\exp (i\Lambda t)\,u_{nm}\,.  \label{heq2}
\end{equation}
The substitution of Eq. (\ref{heq2}) into Eq. (\ref{heq1}) leads to a
time-independent equation for the amplitudes $u_{mn}$. The stationary
solution being known, one can perform the linear-stability analysis around
it in the same way as it has been done for the square lattice \cite
{Eil3,Eil4,JohAub}, assuming a perturbed solution in the form 
\begin{equation}
\psi _{nm}=\exp (i\Lambda t)(u_{nm}+\epsilon w_{nm}),  \label{heq3}
\end{equation}
where $w_{nm}$ is a perturbation with an infinitesimal amplitude $\epsilon $. 
Deriving the leading-order equation for $w_{nm}$, and looking for a
relevant solution to it in the form $w_{nm}=a_{nm}\exp (-i\omega
t)+b_{nm}\exp (i\omega _{nm}^{\star }t)$ (where the eigenfrequency $\omega $
is, generally speaking, complex), one arrives at an eigenvalue problem for
$\{\omega ,\left( a_{nm},b_{nm}^{\star }\right) \}$: 
\begin{eqnarray}
\omega a_{nm} &=&-C\sum_{<n^{\prime },m^{\prime }>}a_{n^{\prime },m^{\prime
}}+kCa_{nm}-2|u_{nm}|^{2}a_{nm}+\Lambda a_{nm}-u_{nm}^{2}b_{nm}^{\star },
\label{heq4} \\
-\omega ^{\star }b_{nm} &=&-C\sum_{<n^{\prime },m^{\prime }>}b_{n^{\prime
},m^{\prime }}+kCb_{nm}-2|u_{nm}|^{2}b_{nm}+\Lambda
b_{nm}-u_{n,m}^{2}a_{nm}^{\star }.  \label{heq5}
\end{eqnarray}

The inclusion of long-range effects into the linear-stability equations is
straightforward. As the long-range coupling is accounted for by a linear
operator acting on the complex field, terms $-{\bf K}\left( h\sqrt{
(n-n^{\prime })^{2}+(m-m^{\prime })^{2}}\right) a_{n^{\prime }m^{\prime }}$
and $-{\bf K}\left( h\sqrt{(n-n^{\prime })^{2}+(m-m^{\prime })^{2}}\right)
b_{n^{\prime }m^{\prime }}$ are to be added to Eqs. (\ref{heq4}) and (\ref
{heq5}), respectively.

\subsection{ILMs in the models with the nearest-neighbor interactions}

Fundamental (single-site-centered) ILM solutions to the stationary equations
were constructed by means of a Newton-type method, adjusted to the
non-square geometry of the TA and HC lattice. For the results presented
herein, we fix the frequency to be $\Lambda =1$ and vary the coupling
constant $C$, as one of the two parameters ($\Lambda $ and $C$) can always
be scaled out from the stationary equations. We started from obvious
single-site solutions (with $|u|=\sqrt{\Lambda }\equiv 1$) at the
anti-continuum limit corresponding to $C=0$ \cite{ma}, and then continued
the solution to finite $C$. Subsequently, the stability analysis was
performed using Eqs. (\ref{heq4})-(\ref{heq5}) for the corresponding lattice.

\begin{figure}[tbp]
\epsfxsize=10cm
\centerline{\epsffile{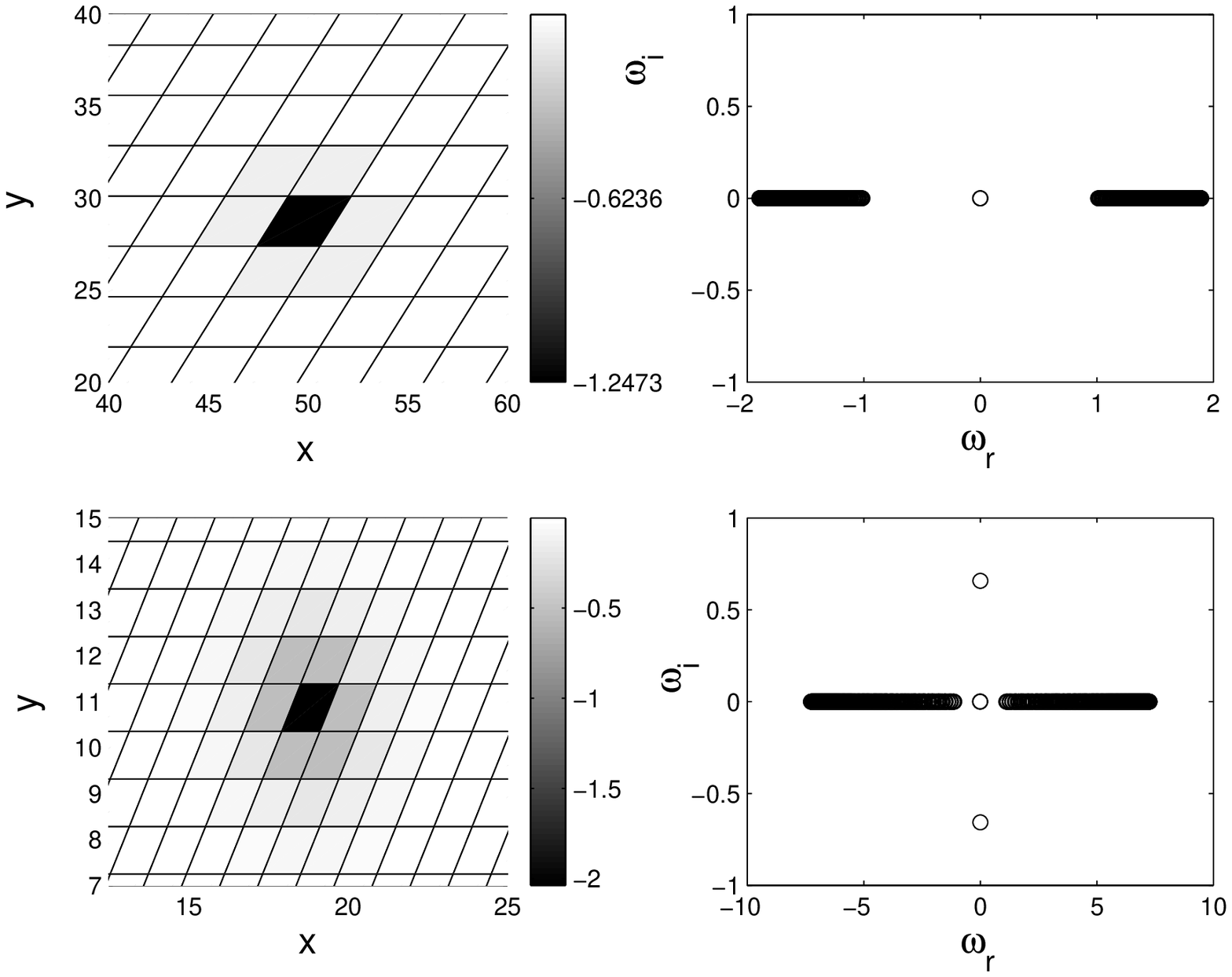}}
\epsfxsize=10cm
\centerline{\epsffile{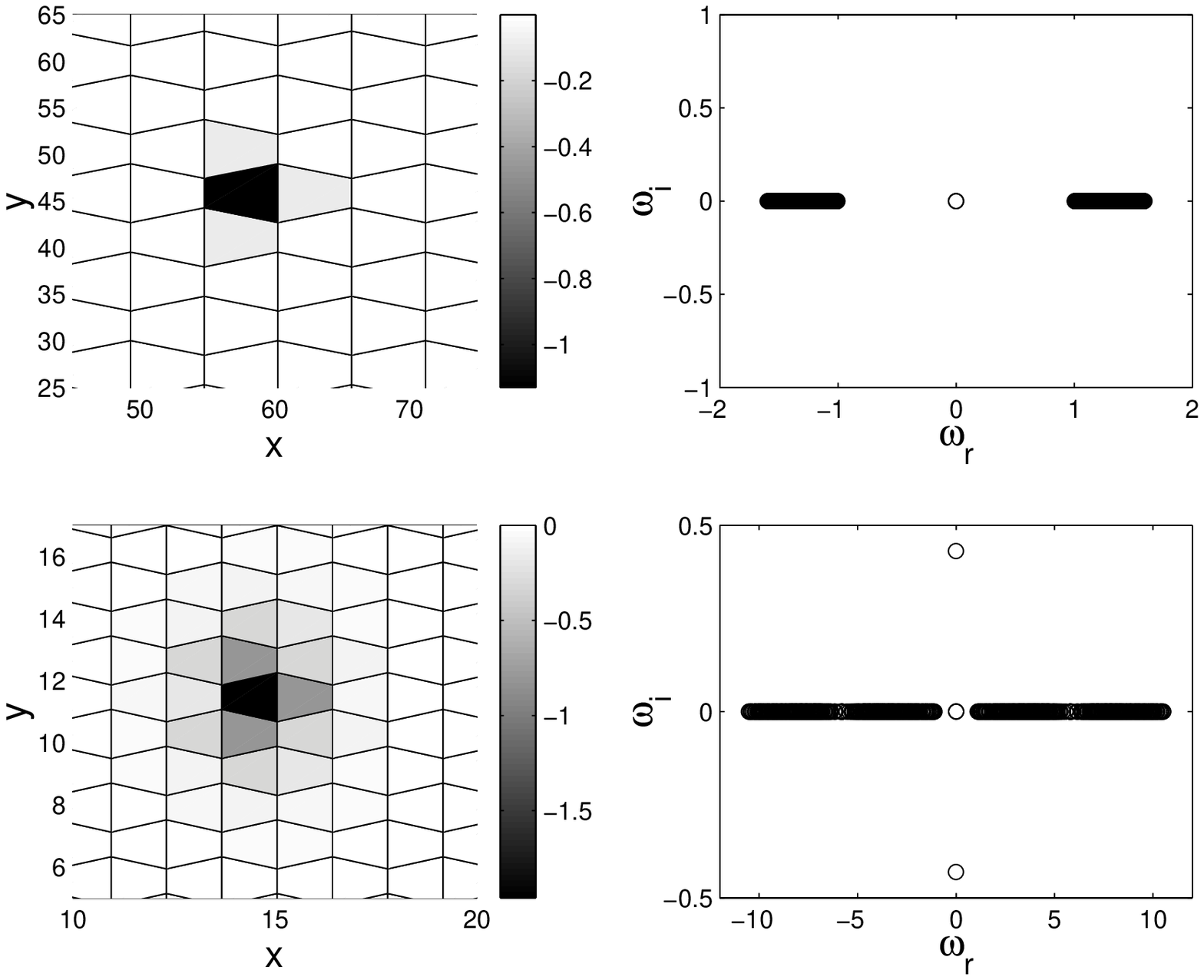}}
\caption{Single-site-centered fundamental ILMs in the triangular and
honeycomb lattices. In the top panel of the figure, the solutions in the
triangular lattice are displayed for two cases: $C=0.1$ in the top row, and $
C=0.7$ in the bottom row. The left subplot in each case shows a checkerboard
contour plot of the solution proper, and the right subplot shows the
spectral plane ($\protect\omega _{r},\protect\omega _{i}$) of the
corresponding eigenfrequencies $\protect\omega $ (the subscripts ${\rm r}$
and ${\rm i}$ stand for the real and imaginary part of the eigenfrequency).
It is seen that the solution is stable for $C=0.1$, and unstable for $C=0.7$. 
Similar results are displayed in the bottom panel of the figure for the
honeycomb lattice. A stable solution is shown for $C=0.1$ in the top row,
and an unstable one for $C=1.6$ in the bottom row. Notice that the contour
plots show a ``negative image'' of the solution. 
The grayscale in all the contour plots presented
in this work is used to denote amplitude.}
\label{hfig4}
\end{figure}

Typical examples of stable and unstable fundamental ILMs found in both the
TA and HC lattices are displayed, by means of contour plots, in Fig. \ref
{hfig4}. The top panel of the figure shows, respectively, stable and
unstable solutions, together with the associated spectral-plane diagrams
(showing the imaginary vs. real parts of the eigenfrequencies), for the TA
lattice with $C=0.1$ (top subplots) and $C=0.7$ (bottom subplots). Stable
and unstable solutions in the HC lattice are shown in the bottom panel of Fig. 
\ref{hfig4} for $C=0.1$ (top subplots) and at $C=1.6$ (bottom subplots).

A natural way to understand the stability of the ILMs is to trace the
evolution and bifurcations of the eigenfrequencies and associated eigenmodes
with the increase of the nearest-neighbor coupling $C$. For the square
lattice, we find, in line with results of Refs. \cite{KRB1,KRB5}, that a
bifurcation generating an internal mode from the edge of the continuous
spectrum (the edge is at $\omega =\Lambda \equiv 1$) in the corresponding
ILM occurs at a critical value $C=0.4486$. As the coupling is further
increased, the pair of the corresponding eigenfrequencies moves towards the
origin of the spectral plane, where they collide and bifurcate into an
unstable pair of imaginary eigenfrequencies at $C=\Lambda $ (i.e., at $
C\equiv 1$ in the present notation), so that the ILM in the square lattice
is unstable for $C>1$.

\begin{figure}[tbp]
\epsfxsize=10cm
\centerline{\epsffile{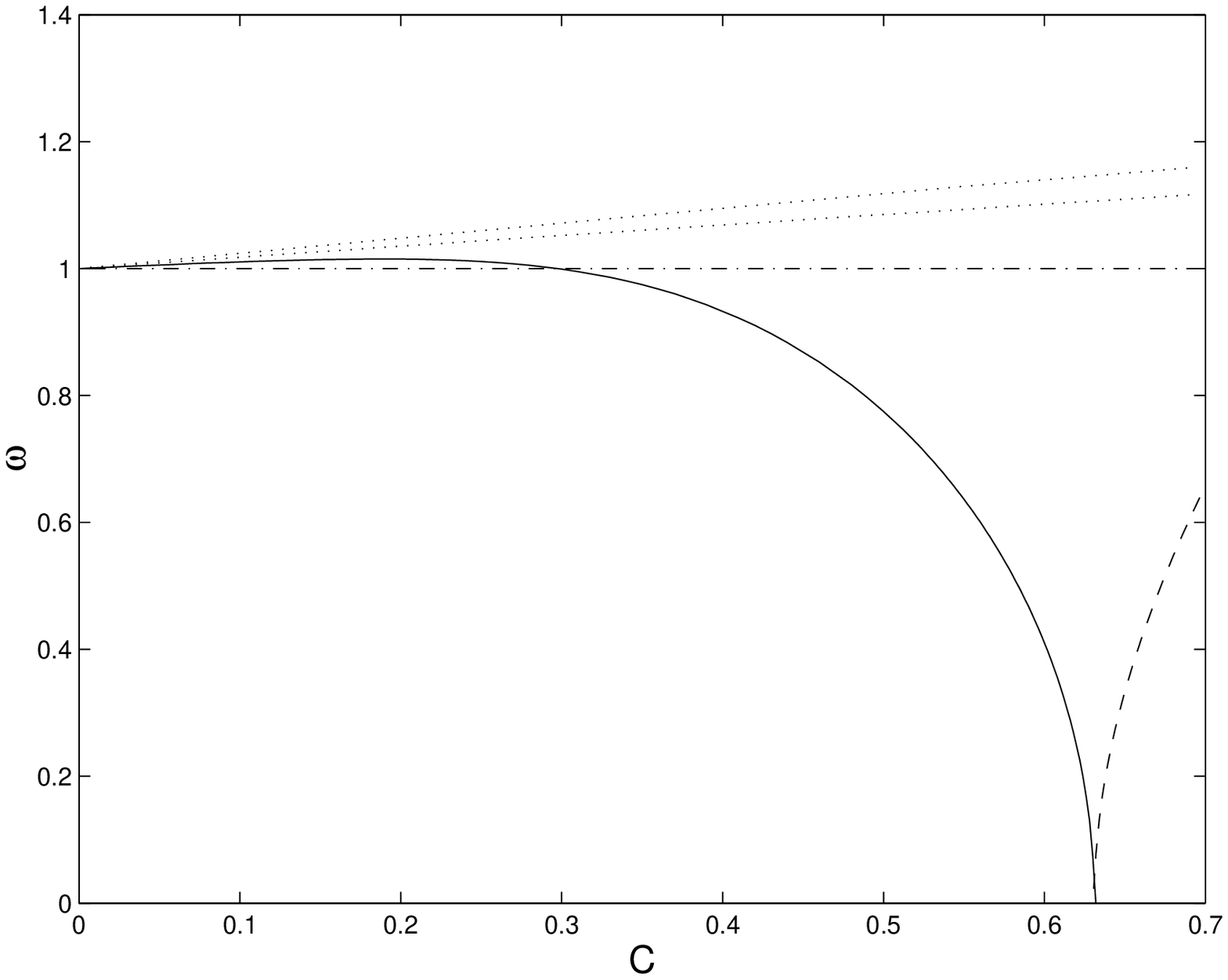}}
\epsfxsize=10cm
\centerline{\epsffile{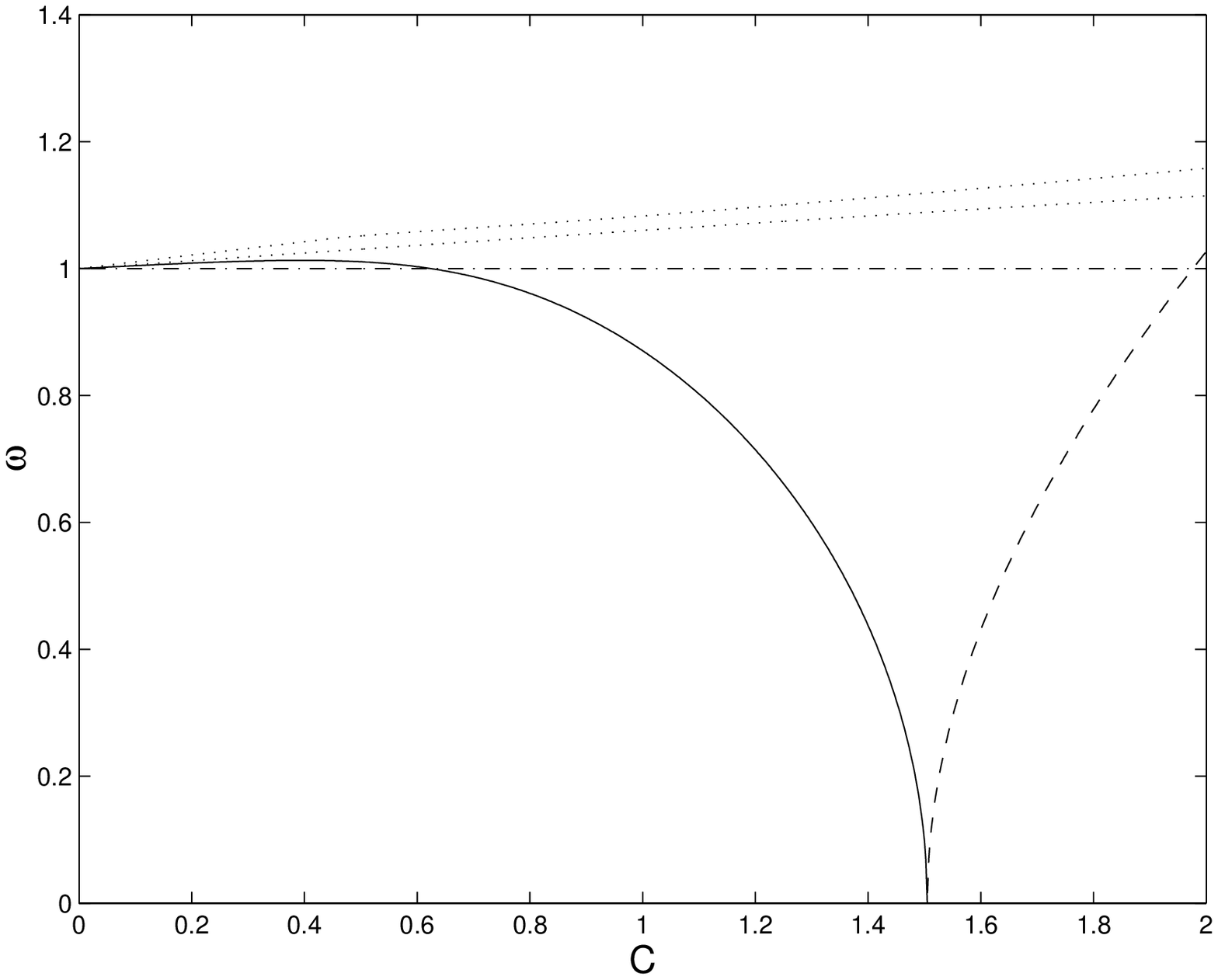}}
\caption{The top panel shows the evolution of the two eigenfrequencies of
the fundamental ILM in the triangular lattice. The eigenvalues bifurcate
from the edge of the continuous spectrum (at $C = 0.2974$) and move towards
the origin, which they hit at $C \approx 0.63$, giving rise to an unstable
pair of imaginary eigenfrequencies. The absolute value of the
eigenfrequencies is shown by the solid line when they are real (stable), and
by the dashed line when they are imaginary (unstable). The dash-dotted line
indicates the edge of the continuous-spectrum band. The bottom panel shows
the same for the ILM in the honeycomb lattice. In this case, the stable
eigenvalues appear at $C \approx 0.6247$, and they become
 unstable imaginary ones, hitting the origin at $C = 1.505$. In both
panels, the two next pairs of eigenvalues (which are always 
found inside the phonon band) are also shown for comparison 
by the dotted lines.}
\label{hfig2}
\end{figure}

In the TA and HC lattices, the scenario is found to be quite similar. For
the former lattice, the bifurcation of the two eigenfrequencies 
from the continuous band edge (which is depicted by the dash-dotted line)
and their trajectory, as they change  from real, i.e., stable (the
solid line) into imaginary, i.e., unstable (the dashed line), are shown in
the top panel of Fig. \ref{hfig2}. The bottom panel shows the same
trajectory for the
HC lattice. The pair of eigenvalues bifurcates at $C=0.2974$ in the TA
lattice, and they reach the origin, giving rise to the instability, at a
point close to $C=0.63$. In the HC lattice, the bifurcation giving rise to
the originally stable eigenvalues occurs at $C=0.6247$; they collide at the
origin and become unstable at $C=1.505$.

\begin{figure}[tbp]
\centerline{\epsffile{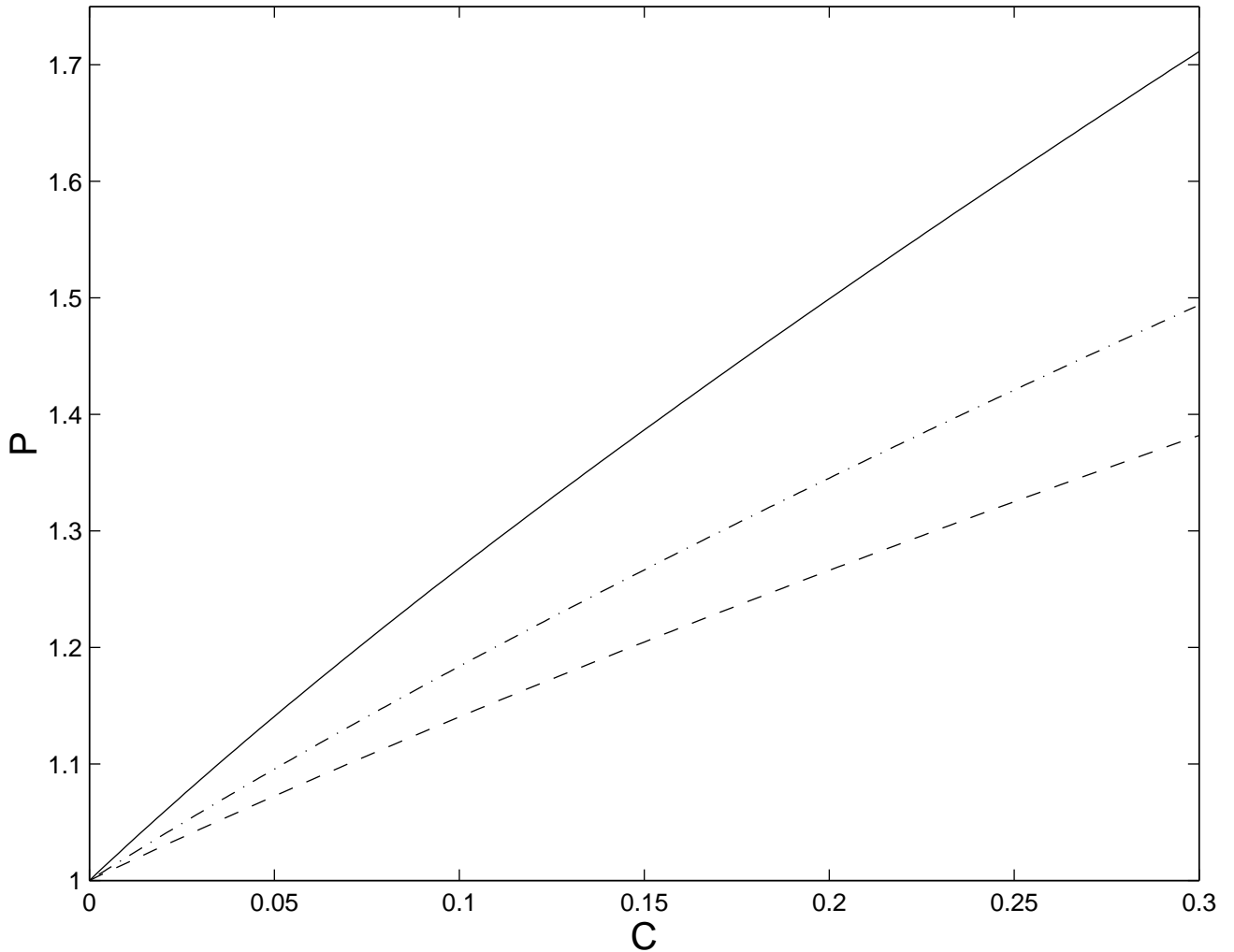}}
\caption{The norm of the fundamental ILM solutions in the TA, square and HC
lattices vs. the coupling constant. The triangular-, square-, and
honeycomb-lattice branches are shown, respectively, by the solid,
dash-dotted, and dashed lines.}
\label{hfig3}
\end{figure}

One can clearly identify the effect of geometry in these results. In
particular, since the instability occurs beyond the critical values of the
coupling, it is the linear interaction between the neighbors that drives it.
Consequently, since the coordination numbers for the different lattices are
ordered as $k_{{\rm triang}}>k_{{\rm square}}>k_{{\rm honey}}$, the
instability thresholds (critical values of the coupling constant) for these
lattices should be ordered conversely, $C_{{\rm triang}}^{{\rm (cr)}}<C_{
{\rm square}}^{{\rm (cr)}}<C_{{\rm honey}}^{{\rm (cr)}}$. A similar
understanding of the effect of the coordination number on the norm of the
solution, $N\equiv \sum_{m,n}\left| u_{mn}\right| ^{2}$, justifies the
results displayed in Fig. \ref{hfig3}: the larger number of neighbors endows
the TA branch (solid line) with a larger norm than the square one
(dash-dotted), which, in turn, has a larger norm than the HC lattice
(dashed).

\subsection{ILMs in the lattices with the long-range interaction}

Recently, a lattice model with a long-range coupling, which is relevant to
magnon-phonon, magnon-libron and exciton-photon interactions, was introduced
in Ref. \cite{gaid}. In the framework of this model, it has been concluded
that the relevant coupling kernel [see Eq.(\ref{heq1})] is 
\begin{equation}
{\bf K}\left( h\sqrt{(n-n^{\prime })^{2}+(m-m^{\prime })^{2}}\right)
=F_{0}K_{0}\left( \alpha h\sqrt{(n-n^{\prime })^{2}+(m-m^{\prime })^{2}}
\right) ,  \label{heq6}
\end{equation}
where $F_{0}$ is an amplitude of the kernel, $\alpha ^{-1}$ measures the
range of the interaction, and $K_{0}$ is the modified Bessel function. We
will use this kernel below.

Inserting the kernel (\ref{heq6}) into Eq. (\ref{heq1}), one can see how the
behavior of the branch is modified as a function of $F_{0}$ for a given
fixed value of the nearest-neighbor coupling $C$. Notice that similar results
(but on a logarithmic scale) will be obtained if $\alpha $ is varied, while $
F_{0}$ is kept fixed, as it was detailed in Ref. \cite{gaid}. Thus, we fix $
\alpha =0.1$ hereafter.

\begin{figure}[tbp]
\centerline{\epsffile{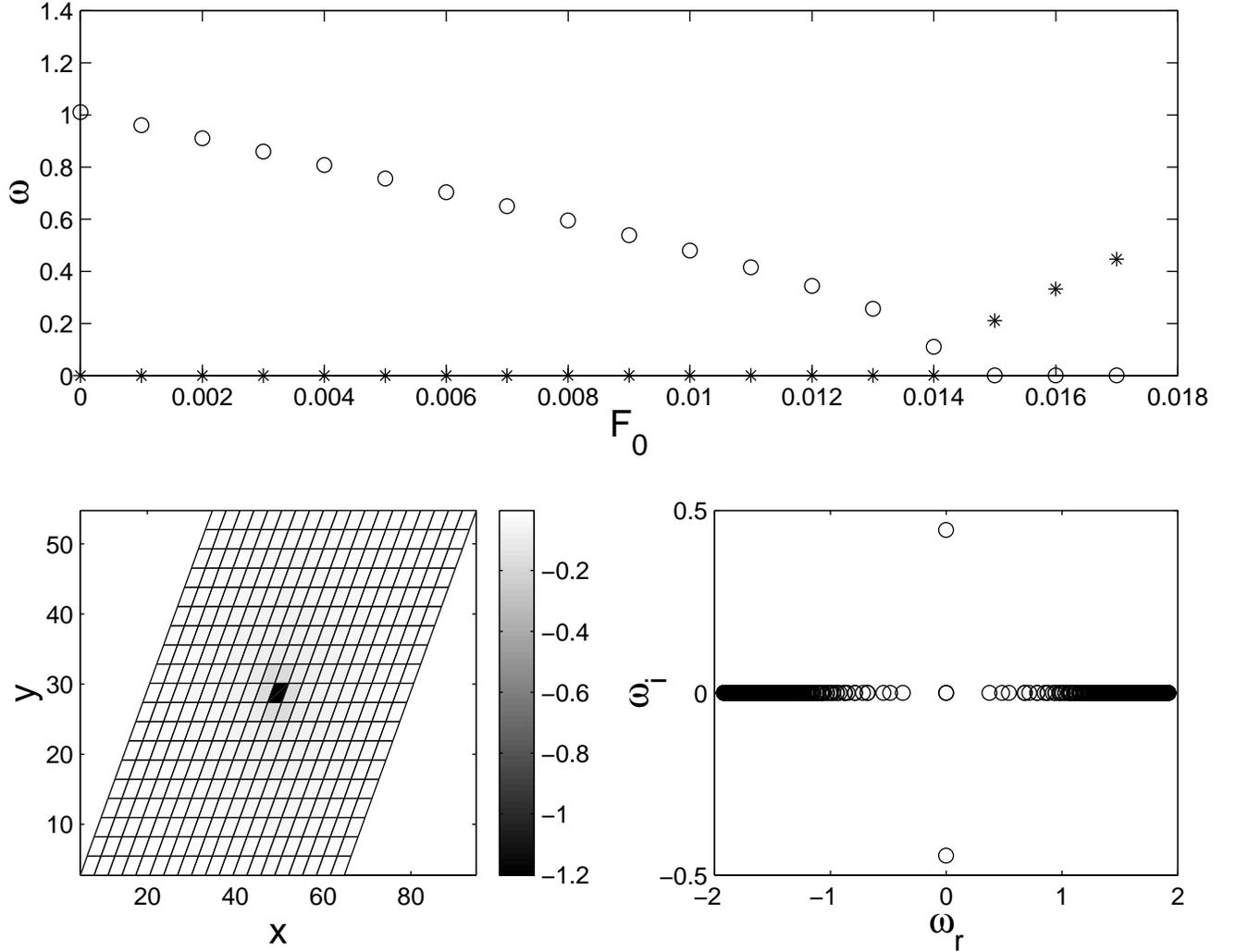}}
\caption{Top panel: the variation of the internal-mode eigenfrequency $
\protect\omega$ for the fundamental ILM vs. the amplitude of the long-range
interaction $F_0$ for $C=0.1$. The stars and circles denote the real and
imaginary parts of the eigenvalue. It can be seen that, while the ILM is
stable at $F_0=0$, an instability sets in at $F_0 = 0.01425$. The left part
of the bottom panel shows the (unstable) ILM configuration, in a
pseudocolor contour plot (of the amplitude), 
and the corresponding spectral plane ($\protect
\omega_r,\protect\omega_i$)) of the eigenmodes is shown in the right part.
These results pertain to the triangular lattice. Once again, the contour
plot shows the ``negative image" of the solution, for clarity.}
\label{hfig5}
\end{figure}

In Fig. \ref{hfig5}, we show the evolution of the numerically found
internal-mode eigenvalues of ILM as a function of $F_{0}$, for fixed $C=0.1$
. It can be observed that the increase of $F_{0}$ leads to an instability
for $F_{0}>0.01425$. The bottom panel shows the configuration and its
internal-eigenmode frequency for $F_{0}=0.015$, when the configuration is
already unstable. Carefully zooming into the ILM's in the case of long-range
interactions (data not shown here),
one can notice a ``tail'' of the ILM, much
longer than the size of the ILM in the case of the nearest-neighbor
interaction, which is a natural consequence of the nonlocal character of the
interaction in the present case. For $F_{0}>0$, we thus conclude that the
long-range interaction ``cooperates'' with the short-range one, lowering the
instability threshold. On the contrary, numerical results for $F_{0}<0$ show
that the onset of the instability is {\em delayed} when the long- and
short-range interactions compete with each other (see also Ref. \cite{gaid}).

\begin{figure}[tbp]
\centerline{\epsffile{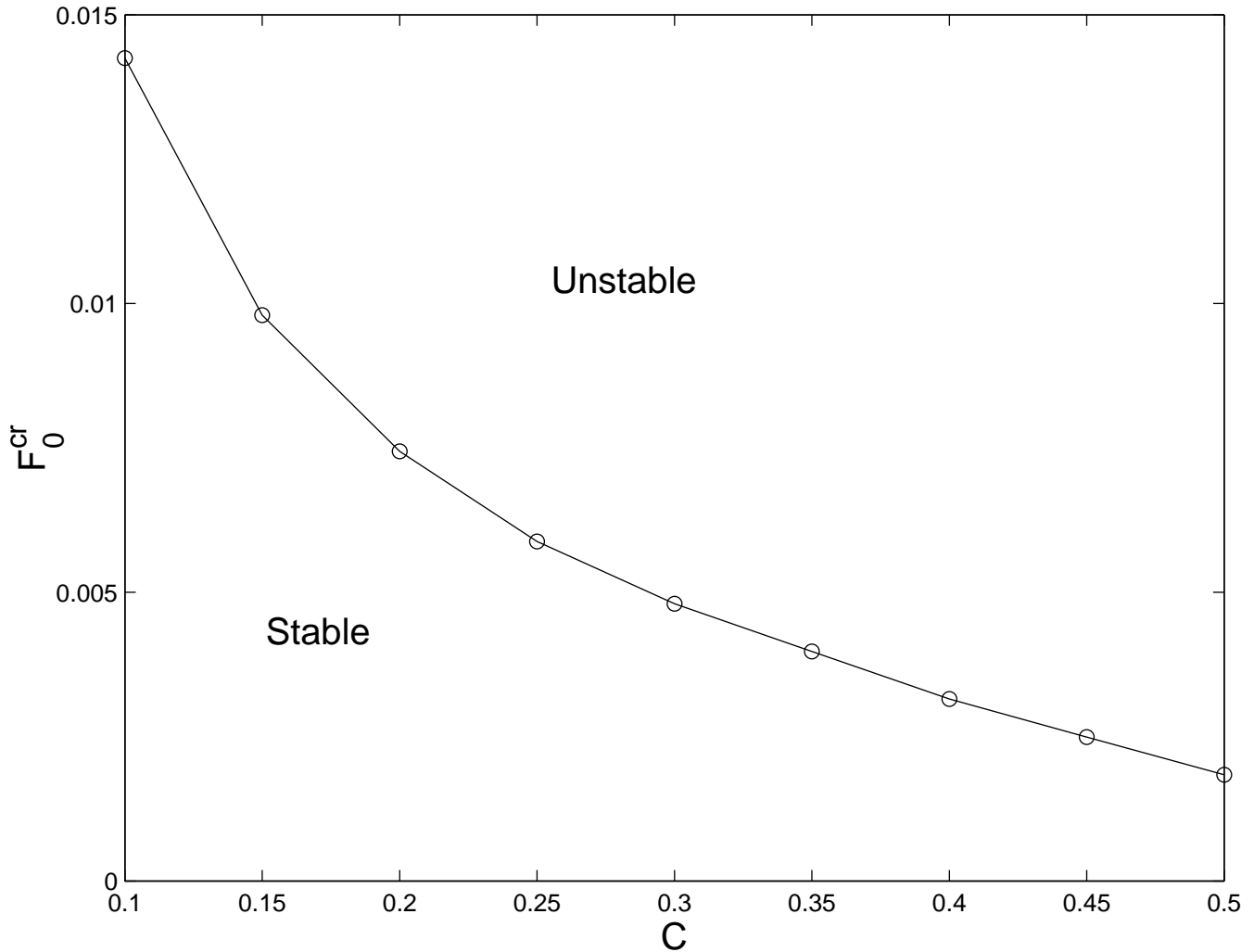}}
\caption{A two-parameter stability diagram for the ILM in the model
combining the long- and short-range interactions. The stability region is
located beneath the curve. The results pertain to the triangular lattice.
For the honeycomb lattice, the stability region is quite similar.}
\label{hfig5a}
\end{figure}

One can extend the above considerations to the case where both $F_{0}$ and $
C $ are varied and construct two-parameter diagrams, separating stability
and instability regions. An example is shown for the TA lattice in Fig. \ref
{hfig5a}. For a fixed $C$, the critical value $\left( F_{0}\right)_{{\rm cr}
} $ was identified, beyond which the ILM is unstable. Thus, in Fig. \ref
{hfig5a} ILM's are stable below the curve and unstable above it.

\section{Multiple-Site ILM's}

We now turn our attention to solutions comprising many sites of the lattice.
In this case too, the solutions are initially constructed in the
anti-continuum limit $C=0$, and then extended through continuation to finite
values of $C$.

\subsection{Twisted localized modes}

Firstly, we examine the so-called twisted localized modes (TLMs), which were
originally introduced, in the context of 1D lattices, in Refs. \cite
{Lad,JohAub2}. Later, they were studied in more detail in Ref. \cite{DKL1},
and their stability was analyzed in Ref. \cite{Kev1}. They were subsequently
used to construct topologically charged 2D solitons (vortices) in the
square-lattice DNLS equation in \cite{DV}.

In the case of the square lattice, and subject to the same normalization as
adopted above, i.e., with $\Lambda =1$, TLMs are found to be stable for $
C<0.125$. If the coupling exceeds this critical value, an oscillatory
instability, which is manifested through a quartet of complex eigenvalues 
\cite{vdm}, arises due to the collision of the TLM's internal mode with the
continuous spectrum (the two have opposite {\it Krein signatures} \cite
{Aub1,JohAub}), as it has been detailed in Ref. \cite{Kev1}. The same
scenario is found to occur in the TA lattice. However, in the latter case
the oscillatory instability sets in at $C\approx 0.1$, and the
destabilization is a result of the collision of the eigenvalues with those
that have (slightly) 
bifurcated from the continuous spectrum (rather than with the edge
of the continuous spectrum at $\omega =1$, as in the square lattice).

\begin{figure}[tbp]
\epsfxsize=10cm
\centerline{\epsffile{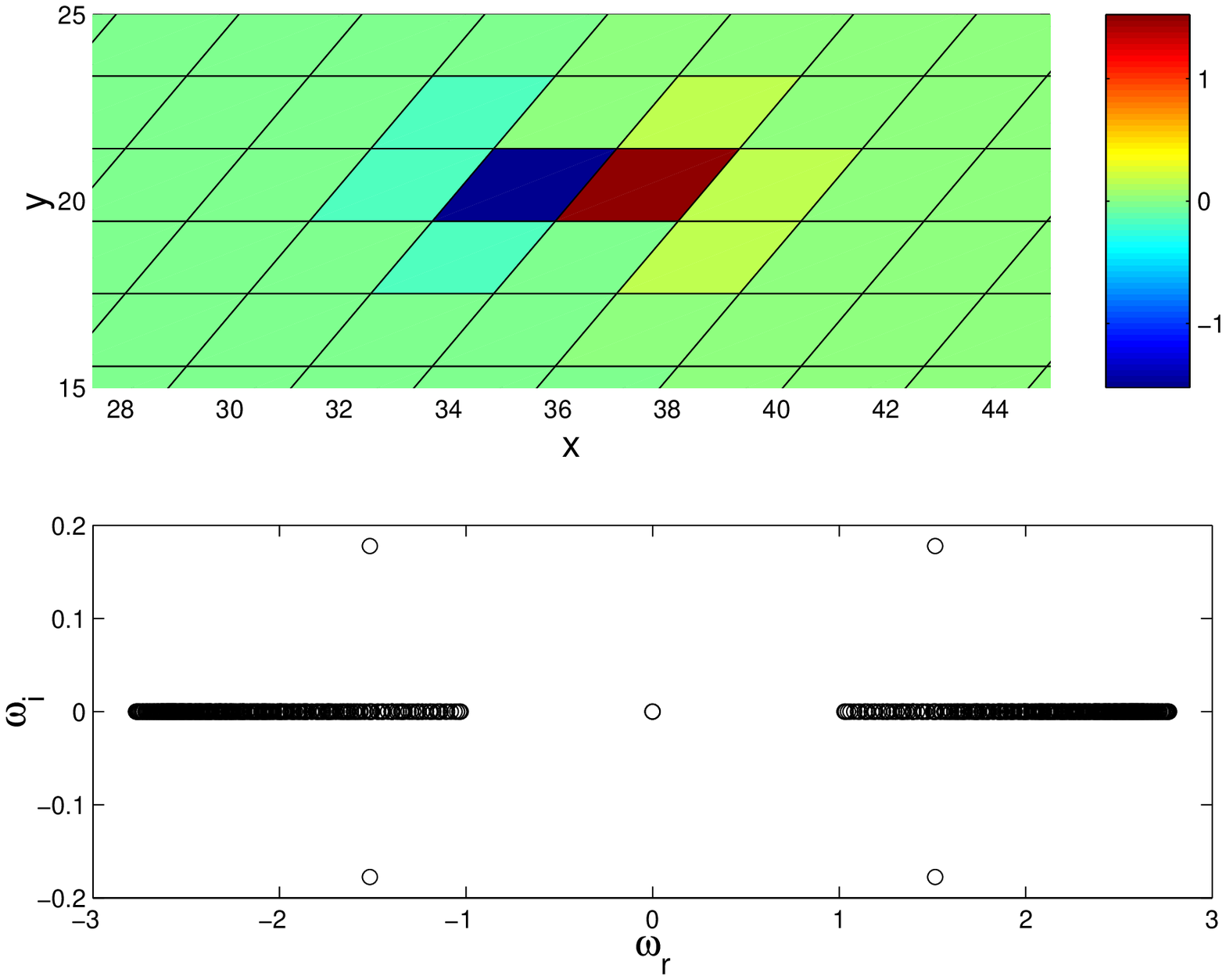}}
\epsfxsize=10cm
\centerline{\epsffile{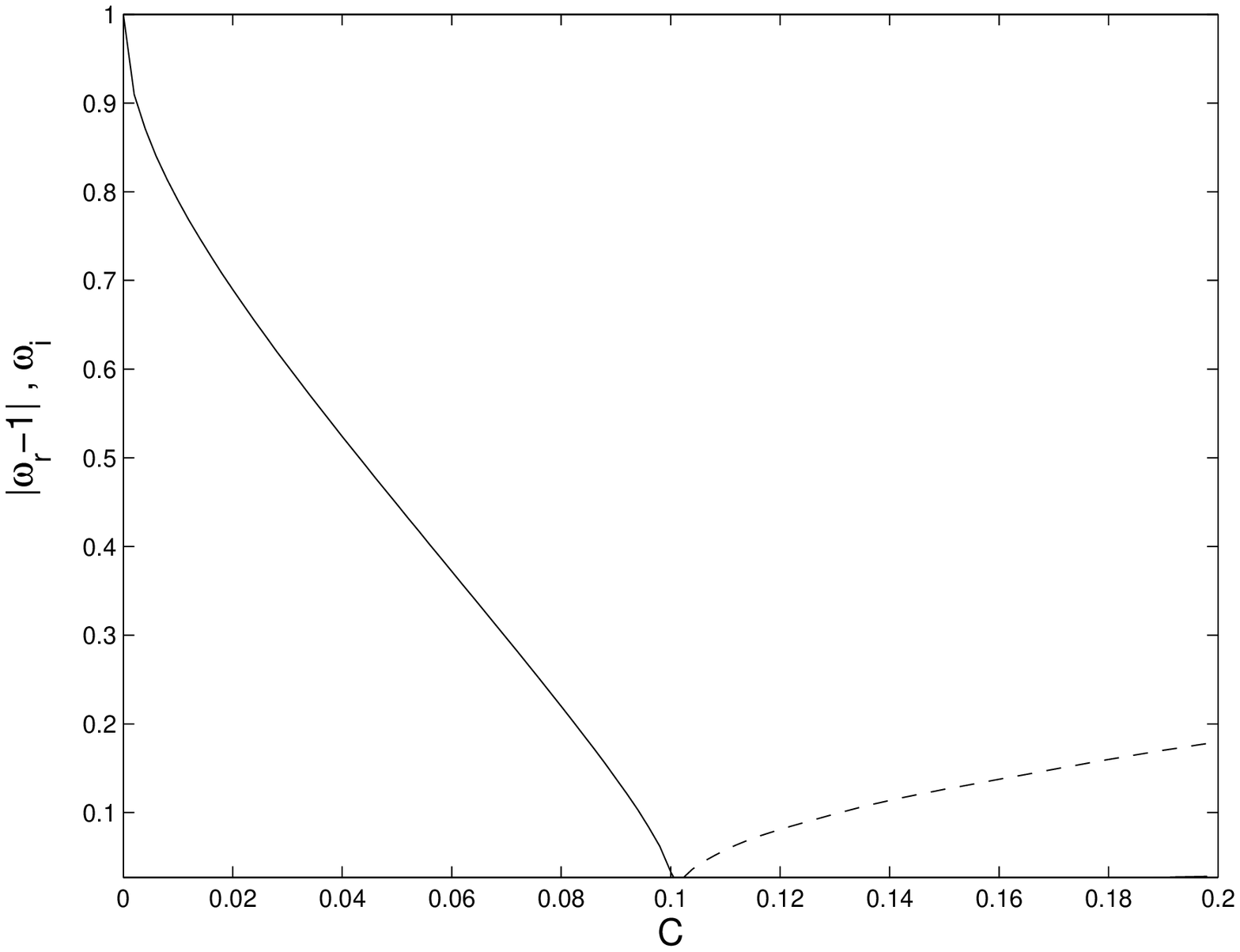}}
\caption{The top panel shows the solution (top subplot) and its stability
(bottom subplot) for a TLM in the triangular lattice at $C=0.2$. One can
readily observe the presence of the oscillatory instability in the
eigenfrequency spectrum. The bottom panel shows the critical eigenfrequency
vs. the coupling constant $C$. The solid line shows the distance of the
eigenfrequency from the band edge of the continuous spectrum. After the
collision, which takes place at $C=0.1$, a quartet of complex eigenvalues
emerges; the absolute value of their imaginary part is shown by the dashed
line.}
\label{hfig6}
\end{figure}

\begin{figure}[tbp]
\epsfxsize=10cm
\centerline{\epsffile{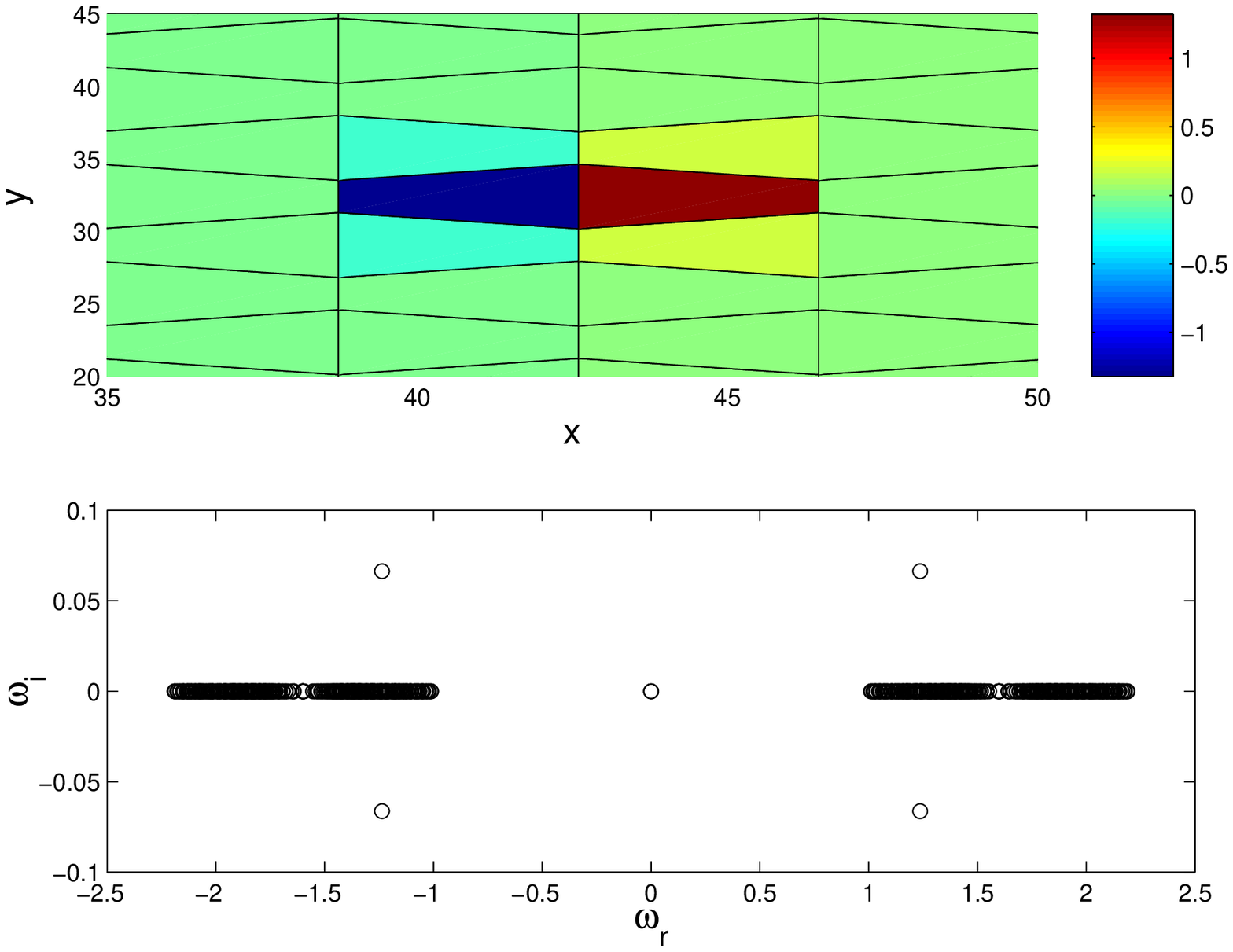}}
\epsfxsize=10cm
\centerline{\epsffile{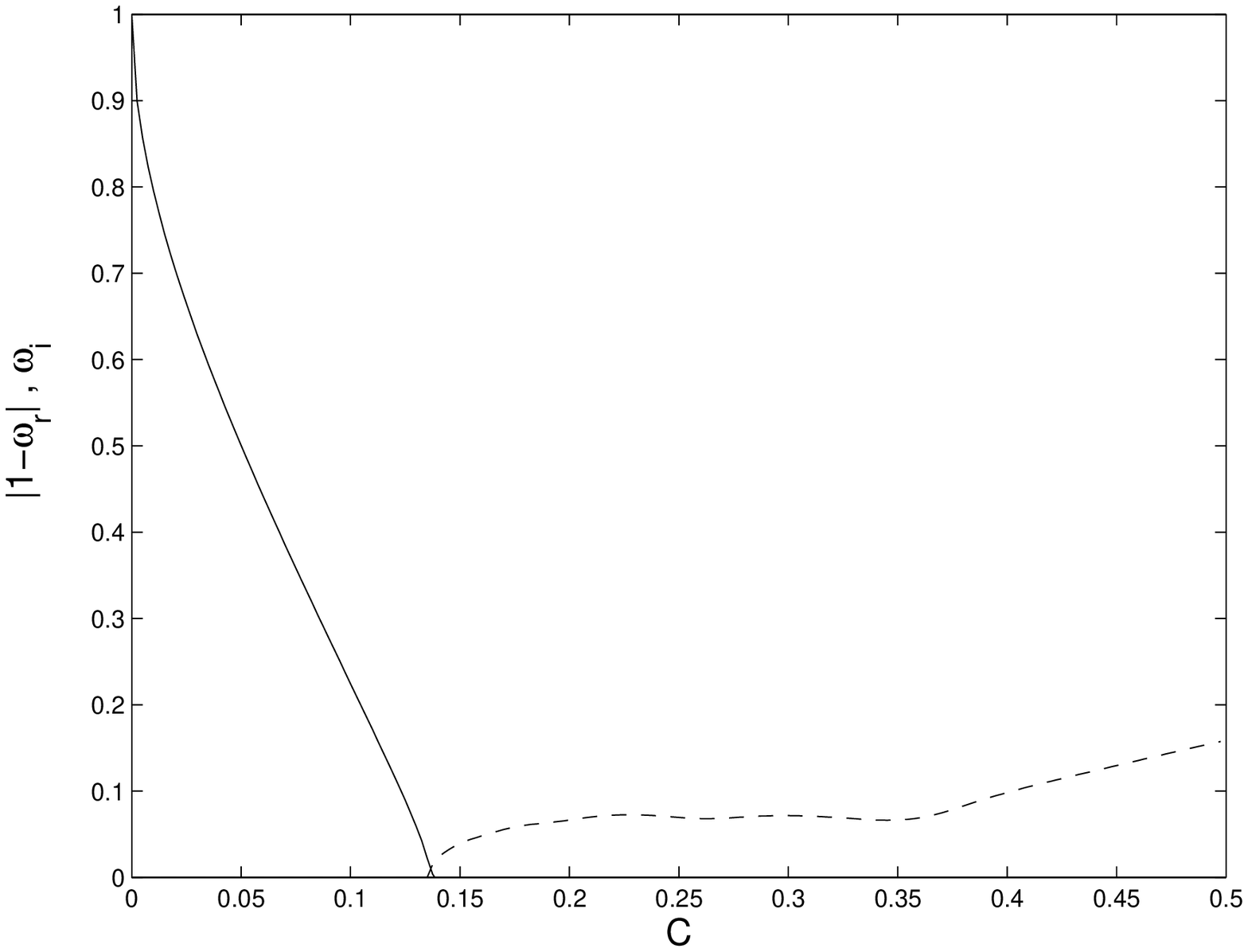}}
\caption{The same as Fig. \ref{hfig6}, but for the honeycomb \ lattice. The
top panel shows the solution at $C=0.2$; the bottom panel shows the distance
of the internal mode eigenfrequency from the band edge (solid line), and the
absolute value of the imaginary part of the eigenvalue quartet for $C>0.1375$
(dashed line).}
\label{hfig6a}
\end{figure}

Similarly, in the HC lattice, the instability of TLMs sets in at $C=0.1375$.
Notice that the instability thresholds follow the same ordering as the ones
discussed in the previous section. This can be justified by a similar line
of arguments as given before. The TLM in the TA lattice (and its stability)
is displayed in the top panel of Fig. \ref{hfig6} for $C=0.2$, which
exceeds the instability threshold. The bottom panel of the figure shows the
variation (as a function of the coupling $C$)
 of the critical eigenfrequency. The real stable eigenfrequency,
and the imaginary part of the unstable ones, after the threshold has been
crossed, are shown, respectively, by solid and dashed lines. Figure \ref
{hfig6a} displays analogous results for the HC lattice. The solution is
shown at $C=0.2$ in the top panel.

It should be remarked that, in the 2D lattice, the TLMs are solutions
carrying vorticity (topological charge) $S=1$ \cite{DV} (although they are
different from vortices proper, see below). The simplest way to see this is
by recognizing that TLM configurations emulate the continuum-limit
expression $\cos \theta $, where $\theta $ is the angular coordinate in the
2D plane, i.e., the real part of $\exp (i\theta )$, the latter expression
carrying vorticity $1$. It should also be added that, after the onset of the
oscillatory instability, TLM solutions have been found to transform
themselves into the fundamental (single-site-centered) ILM configurations,
which is possible as the topological charge is not a dynamical invariant in
lattices \cite{DV,chi2}.

\subsection{Hexagonal and triangular vortex solitons}

\begin{figure}[tbp]
%\centerline{\epsfxsize=7cm}
\centerline{\epsffile{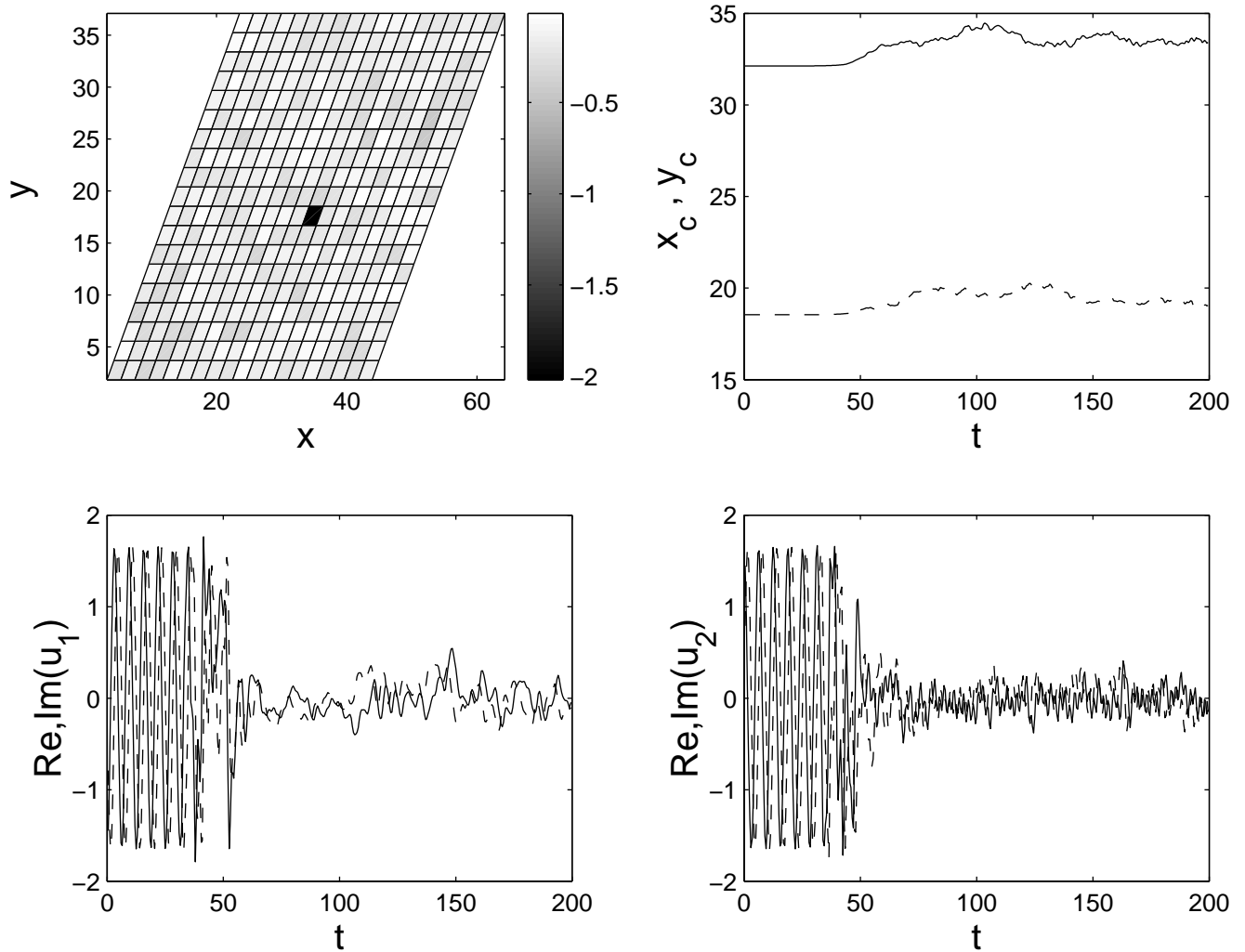}}
\caption{The time evolution of the unstable solution from the previous
figure (for $C=0.218$) is shown here for the triangular lattice. The top
left panel shows the solution at $t=200$. The ``negative image" 
of the solution is shown once again for clarity. 
The top right panel shows the time evolution of the
center-of-mass coordinates of the soliton, defined as $x_c \equiv \sum_{n,m}
n |u_{n,m}|^2/\sum_{n,m} |u_{n,m}|^2$, $y_c \equiv \sum_{n,m} m
|u_{n,m}|^2/\sum_{n,m} |u_{n,m}|^2$. The two bottom panels show the
evolution of the real (solid line) and imaginary (dashed line) parts of the
lattice field at two sites closest to the soliton's center (subscripts $1 $
and $2$ pertain, respectively, to the sites $n=10,m=11$ and $n=10, m=9$). A
clear conclusion is that the instability sets in at $t \approx 50$, and it
eventually results in the transformation of the hexagonal ILM into a stable
fundamental ILM localized around a single lattice site.}
\label{hfig8}
\end{figure}

Going beyond TLMs, it is appropriate to consider possible lattice solitons
which conform to the symmetry of the TA or HC lattice. In fact, these are
the most specific dynamical modes supported by the lattices. An example of
this sort in the TA lattice are ``hexagonal'' ILMs shown in Fig. \ref{hfig8}
. The top panel of the figure shows the profile of these modes in the
anti-continuum limit, and the bottom panel displays two actual examples of
these modes. The top subplot shows the hexagonal ILM for $C=0.038$, when it
is stable, while the bottom subplot shows the mode at $C=0.218$, after the
onset of three distinct oscillatory instabilities. The first and second
instabilities set in at $C\approx 0.064$ and $C\approx 0.084$ respectively,
while the final eigenvalue quartet appears at $C\approx 0.184$.

\begin{figure}[tbp]
\epsfxsize=10cm
\centerline{\epsffile{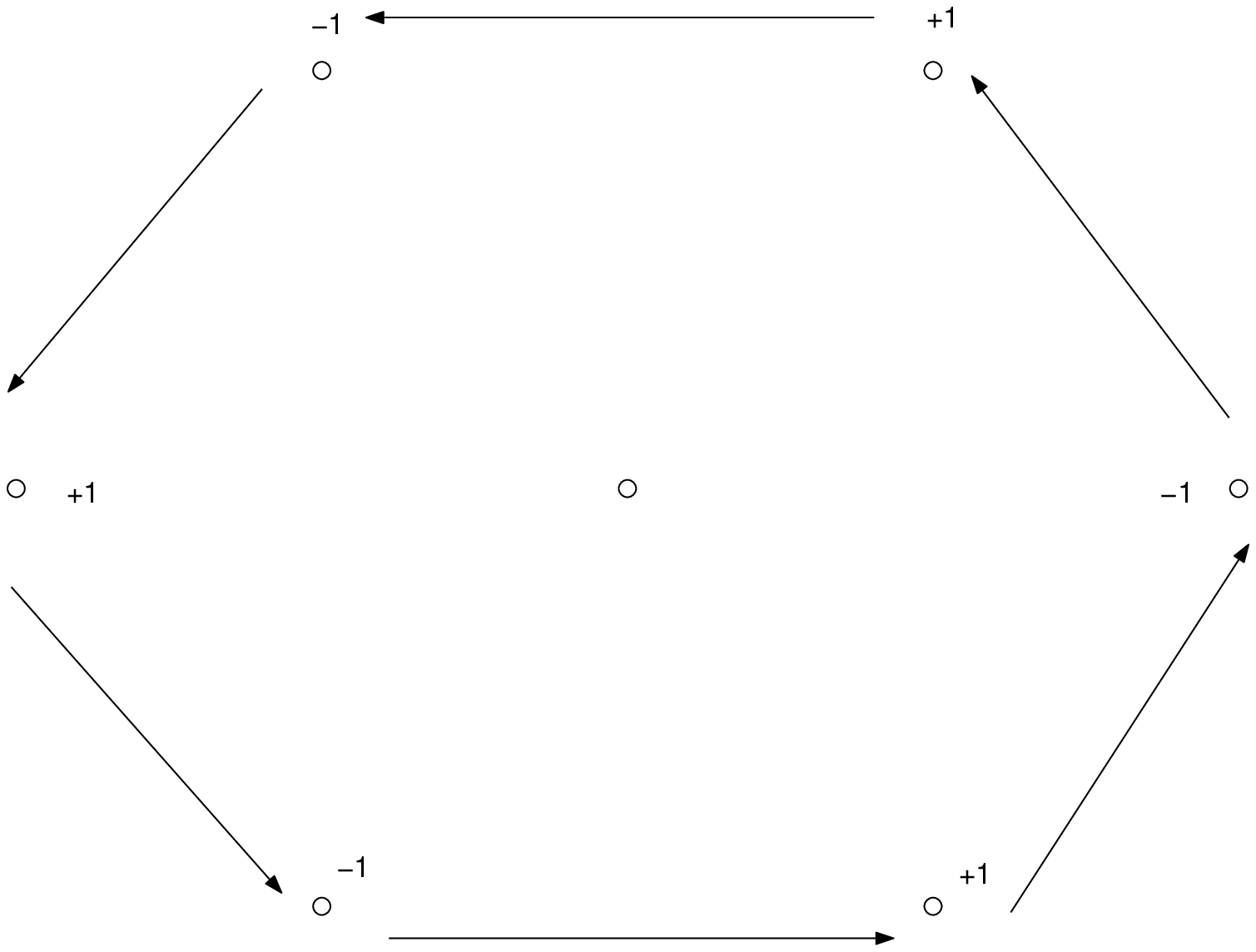}}
\epsfxsize=10cm
\centerline{\epsffile{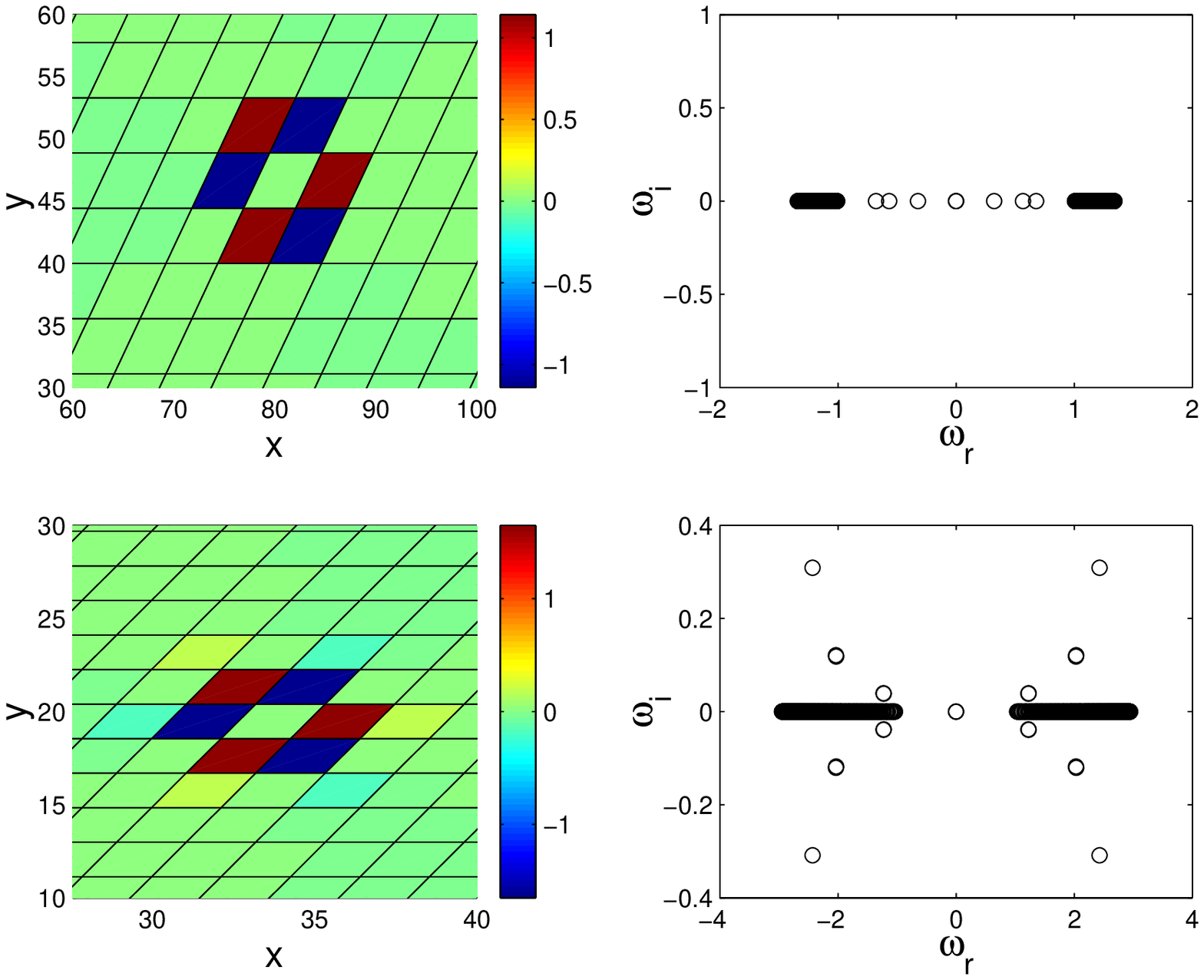}}
\caption{The top panel shows the anti-continuum-limit profile of the
topologically charged hexagonal ILM in the triangular lattice (and a contour
around the solution; each line in the contour represents a phase change by $
\protect\pi$). The bottom panel shows two examples of such a soliton. The
solution is in each case shown in a pseudocolor contour plot on the left,
while its eigenfrequency spectrum is shown on the right. The top subplot of
the bottom panel corresponds to the stable soliton at $C=0.038$, while the
bottom subplot represents an unstable one at $C=0.218$. In the latter case,
three (in terms of the geometric multiplicity; the algebraic multiplicity is
five) quartets of eigenvalues have become unstable.}
\label{hfig7}
\end{figure}

Measuring the topological charge of this solution around the contour in the
top panel of Fig. \ref{hfig7}, we find (since each jump from $+1$ to $-1$
can be identified as a $\pi $ phase change) that the whole solution has a
total phase change of $6\pi $, hence its topological charge (vorticity, or
``spin'') is $S=3$. Then, the presence of three oscillatory instabilities
agrees with a recent conjecture \cite{chi2}, which states that the number of
negative-Krein-sign eigenvalues (and hence the number of potential
oscillatory instabilities) coincides with the topological charge of the 2D
lattice soliton. However, if one examines more carefully the stability
picture, one finds that, due to the symmetry of the solution, two of these
eigenvalues have multiplicity $2$. Hence, the conjecture needs to be
refined, to
take into regard the potential presence of symmetries. The thus revised
conjecture states that the topological charge of the solution should be
equal to the geometric (but not necessarily algebraic) multiplicity of the
eigenvalues with negative Krein signature.

It is natural to ask then to what configuration this hexagonal ILM will
relax once it becomes unstable. To address the issue, we performed direct
numerical simulations for $C=0.218$. Results are shown in our subplots of
Fig. \ref{hfig8}. In particular, the top left panel shows the solution at $
t=200$ (the configuration at $t=0$ was the unstable hexagonal ILM). It can
clearly be observed that the instability that sets in around $t\approx 50$
(according to the other three subplots) transforms the hexagonal vortex into
a fundamental (zero-vorticity) ILM; recall that such an outcome of the
instability development is possible because the vorticity is {\em not}
conserved in lattice systems \cite{DV}.

\begin{figure}[tbp]
\centerline{\epsffile{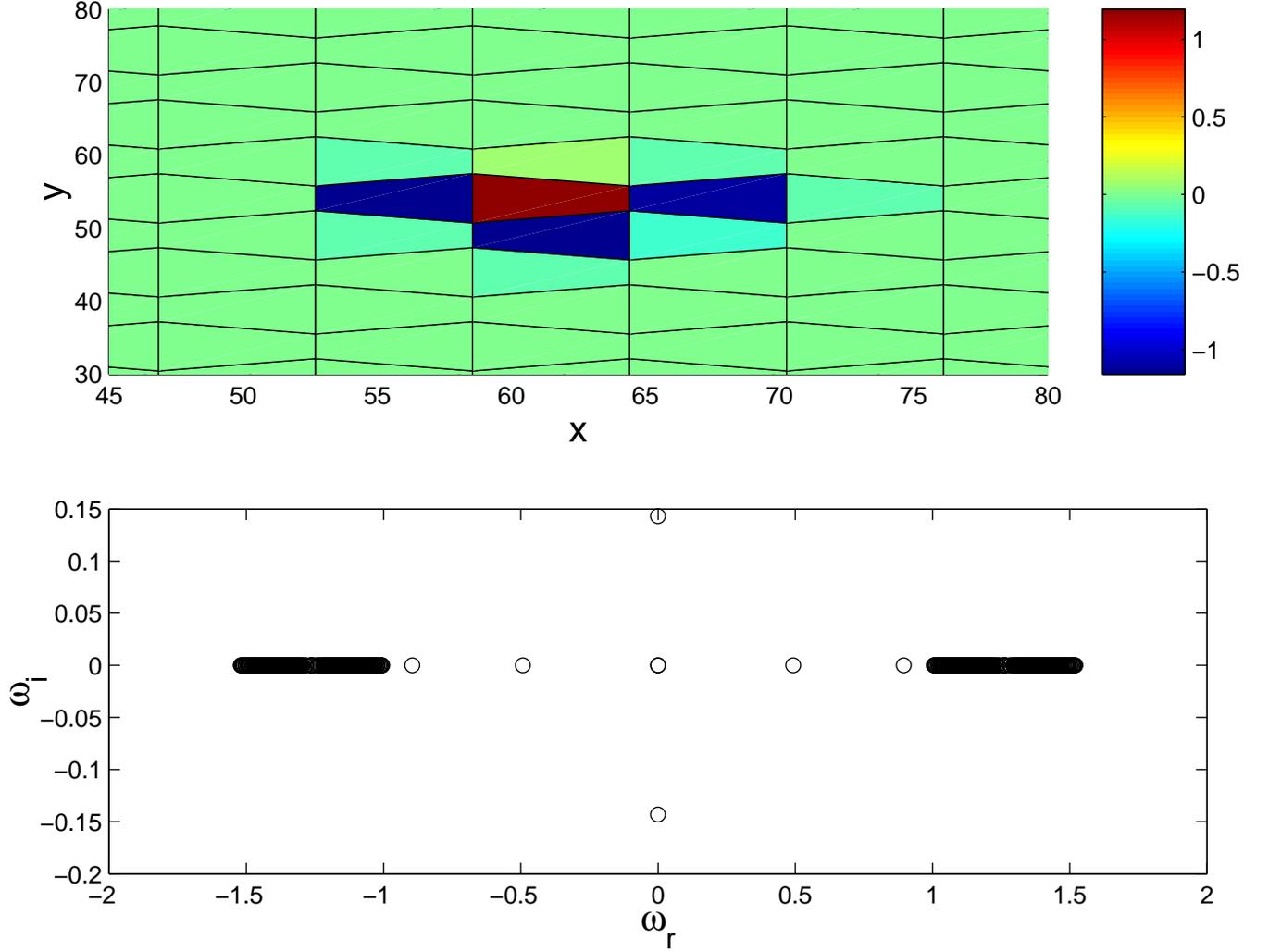}}
\caption{A triangular ILM solution in the honeycomb lattice, and the
corresponding spectral plane, are shown for $C=0.09$. Solutions of this type
were found to be unstable for {\em all} values of $C$.}
\label{hfig9}
\end{figure}

For the HC lattice, a vortex soliton of a triangular form was found, see an
example in Fig. \ref{hfig9} for $C=0.09$. We have found that this solution
is unstable for {\em all values} of $C$.

\begin{figure}[tbp]
\epsfxsize=10cm
\centerline{\epsffile{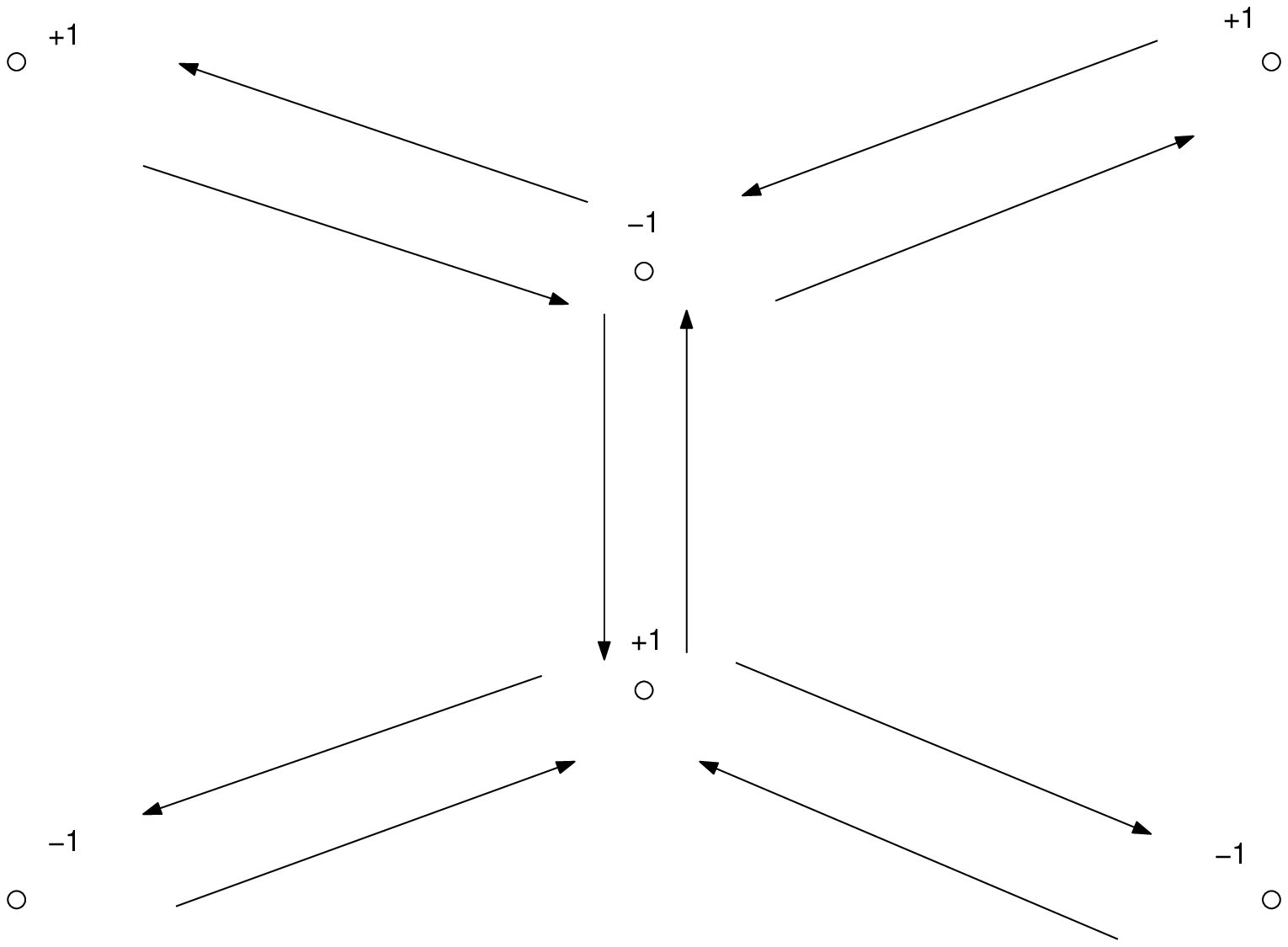}}
\epsfxsize=10cm
\centerline{\epsffile{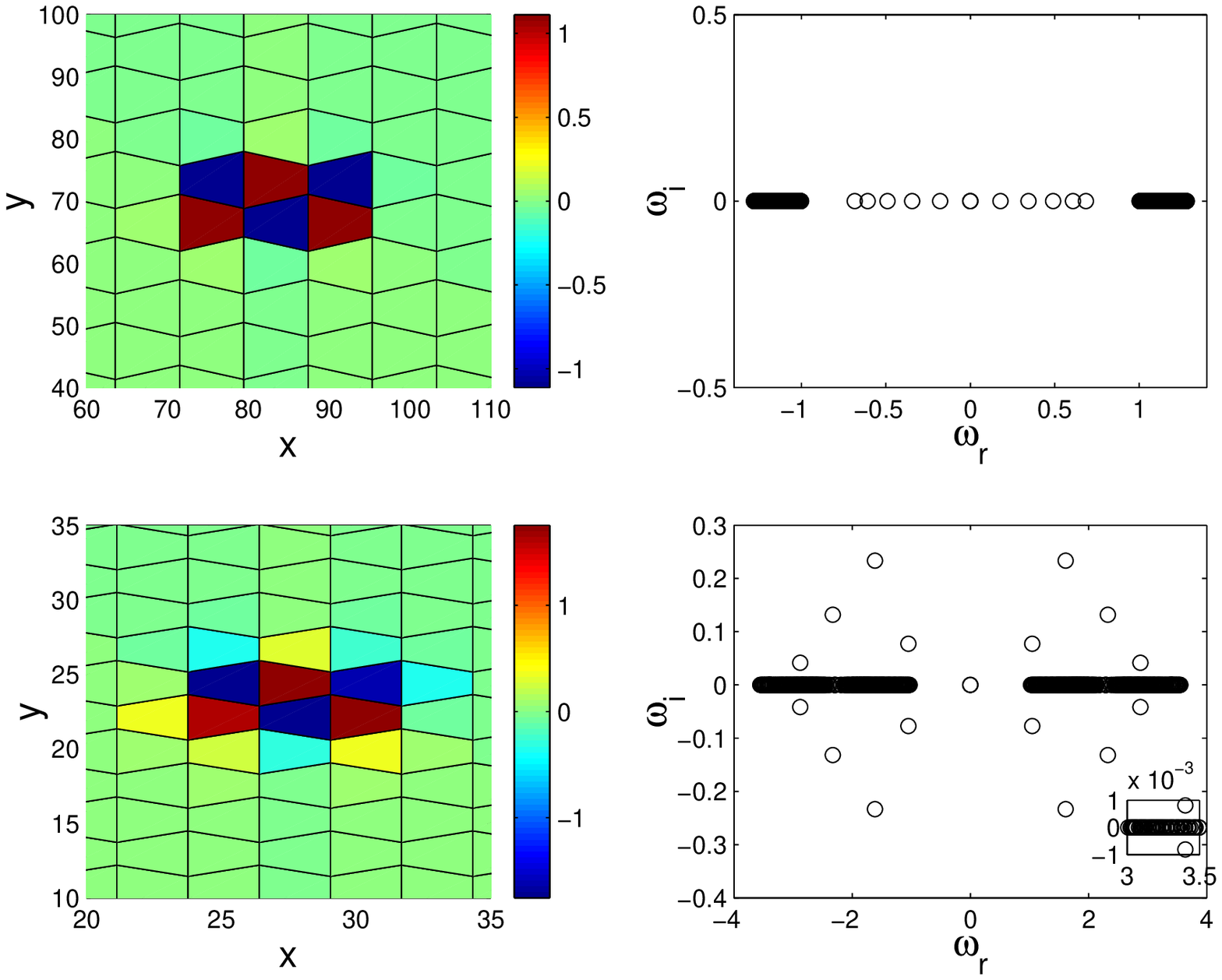}}
\caption{The top panel shows the profile of the honeycomb-shaped ILM in the
anti-continuum limit in the honeycomb lattice, together with a contour
around the solution. Each line in the contour represents a phase change by $
\protect\pi$, adding up to $10 \protect\pi$, hence the solution has
vorticity $S=5$. An actual solution is shown in the bottom panel for 
$C=0.0475$, 
when it is linearly stable, having five negative-Krein-signature internal
modes (top subplot). The bottom subplot of the bottom panel pertains to the
case $C=0.43$, where all five oscillatory instabilities have developed. Four
quartets of eigenfrequencies are clearly discernible. The fifth quartet is
shown in the inset of the corresponding spectral-plane plot.}
\label{hfig10}
\end{figure}

Another vortex soliton, with a honeycomb shape, was also found in the HC
lattice, see Fig. \ref{hfig10}. This one is {\em stable} at a sufficiently
weak coupling. If a contour is drawn around this solution (shown in the top
panel of the figure for the anti-continuum limit), the net phase change is
found to be $10\pi $, hence the corresponding topological charge is $S=5$.
In accordance with the conjecture mentioned above, when the solution is
stable, we find five internal modes with negative Krein signature. These
modes eventually lead, as the coupling is increased, to five oscillatory
instabilities. In the bottom panel of Fig. \ref{hfig10}, the top subplot shows
the case with $C=0.0475$, when the honeycomb-shaped vortex soliton in the HC
lattice is linearly stable. The bottom subplot features the presence of five
eigenvalue quartets in the case $C=0.43$, when all five oscillatory
instabilities have been activated. The first instability occurs at $C=0.085$, 
the second at $0.1$, the third at $0.1375$, the fourth at $0.2025$, and
the fifth sets in at $C=0.4025$.

\begin{figure}[tbp]
\centerline{\epsffile{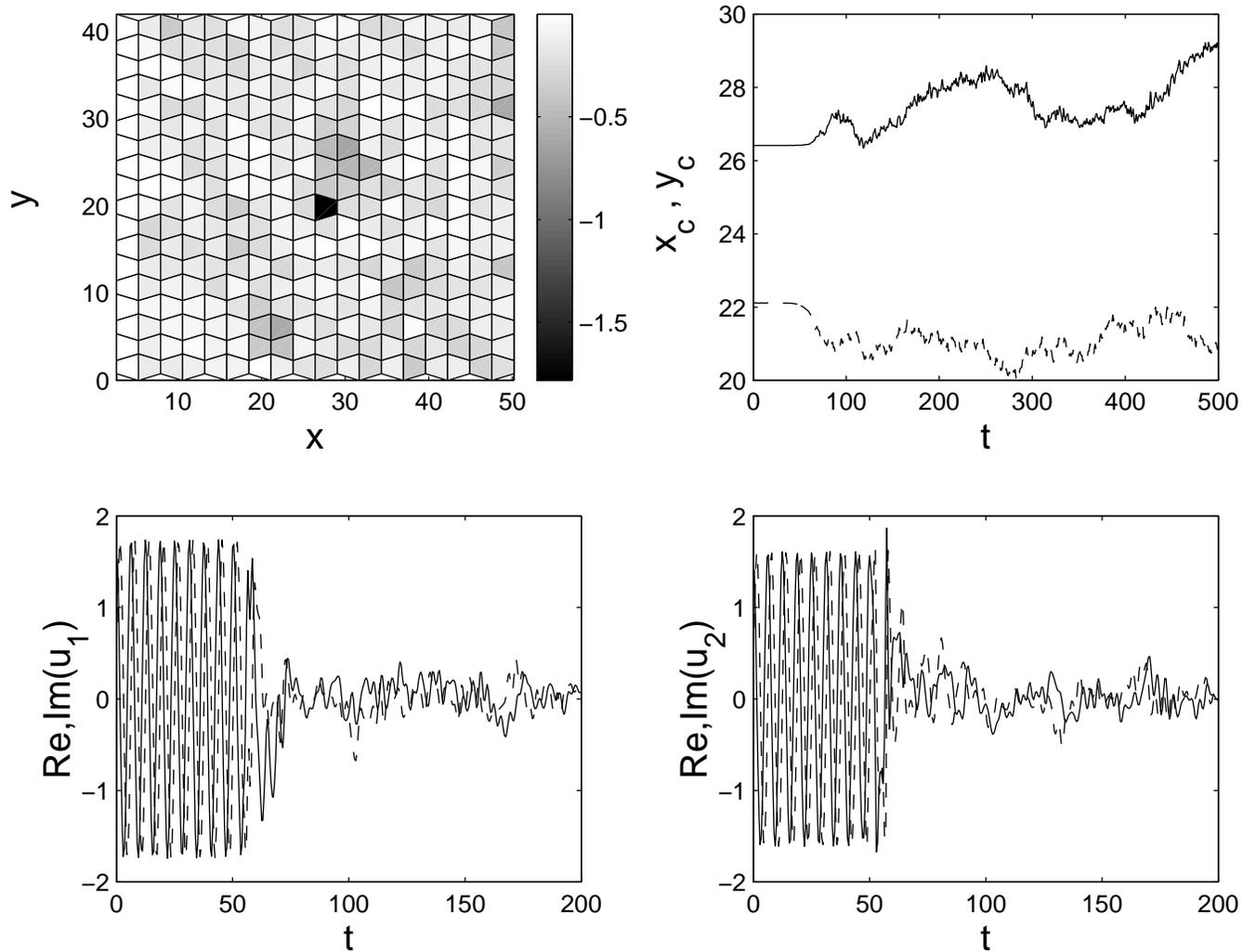}}
\caption{The evolution of the unstable honeycomb-shaped ILM configuration
for $C=0.43$. The panels have the same meaning as in Fig. \ref{hfig8}. The
lattice-field configuration is shown in the top left panel for $t=500$. One
predominant single-site pulse is present in the configuration.
%; a
%smaller-size pulse will eventually decay into phonons. 
The other three
panels show the time evolution of the center-of-mass coordinates and of the
real and imaginary parts of the field at specific lattice sites (as in Fig. 
\ref{hfig8}), clearly indicating that the instability sets in at $t \approx
60$.}
\label{hfig11}
\end{figure}

To illustrate the result of the development of the instability of the latter
vortex in the case in which it is unstable, we have again resorted to direct
numerical integration of Eq. (\ref{heq1}). An example is shown in Fig. \ref
{hfig11} for $C=0.43$. At $t\approx 500$, only one main pulse is sustained.
%; a smaller pulse is also seen, that has already lost a part of its energy and
%will eventually decompose into phonon radiation. 
Hence, in this case too,
the multiply charged topological soliton is transformed, through the
instability, into the stable (for this value of $C$) fundamental
zero-vorticity ILM.

\section{Conclusion}

In this work, we have studied a paradigm nonlinear lattice dynamical model,
namely the DNLS equation, in two spatial dimensions for non-square lattices.
The triangular and honeycomb networks were considered, as the most important
examples of Bravais and non-Bravais 2D lattices, which are relevant to
chemical and optical applications.

In the case of nearest-neighbor interactions, it was found that the
instability thresholds for the fundamental solution centered at a single
lattice site depend on the coordination number. The instability appears in
the triangle lattice at a smaller value of the coupling constant than in the
square lattice, while the opposite is true for the honeycomb lattice. The
effect of the long-range interactions was also examined in this context. It
was found that these interactions accelerate or delay the onset of the
instability if they have the same sign as the nearest-neighbor coupling, or
the opposite sign. Diagrams in the two-parameter plane were constructed,
identifying regions of stability and instability in the presence of both the
short- and long-range coupling.

More complicated lattice solitons, which essentially extend to several
lattice sites, were also examined. A prototypical example of the extended
solitons are twisted modes, for which the phenomenology was found to be
similar to that in the square lattice, but with, once again, appropriately
shifted thresholds. We have also examined solutions with a higher
topological charge, which play the role of fundamental vortices in the
triangular and honeycomb lattices, their vorticity being, respectively, $S=3$
and $S=5$. Stability of these vortices was studied in
detail. When instabilities occurred, their outcome was examined by means of
direct time integration, showing the transformation into a simple
fundamental soliton with zero vorticity.

Further steps in the study of localized modes in these nonlinear lattices
may address traveling discrete solitons, as well as generalization to the
three-dimensional case. In terms of applications, a relevant object are
nonlinear photonic band-gap crystals based on non-square lattices.

The authors are grateful to J.C. Eilbeck for a number of stimulating
discussions.

\end{document}